\documentclass[prb,showpacs,showkeys]{revtex4}
\usepackage{amsfonts}
\usepackage{amsmath}
\usepackage{amssymb}
\usepackage{graphicx}

\begin{document}

\title{Point defects and clustering in uranium dioxide by LSDA+U calculations}
\author{Hua Y. Geng}
\affiliation{Department of Quantum Engineering and Systems Science, The University of Tokyo, Hongo 7-3-1, Tokyo 113-8656,
Japan}
\author{Ying Chen}
\affiliation{Department of Quantum Engineering and Systems Science, The University of Tokyo, Hongo 7-3-1, Tokyo 113-8656,
Japan}
\author{Yasunori Kaneta}
\affiliation{Department of Quantum Engineering and Systems Science, The University of Tokyo, Hongo 7-3-1, Tokyo 113-8656,
Japan}
\author{Misako Iwasawa}
\affiliation{Materials Science Research Laboratory, Central Research Institute
of Electric Power Industry, Tokyo 201-8511, Japan}
\author{Toshiharu Ohnuma}
\affiliation{Materials Science Research Laboratory, Central Research Institute
of Electric Power Industry, Tokyo 201-8511, Japan}
\author{Motoyasu Kinoshita}
\affiliation{Nuclear Technology Research Laboratory, Central Research Institute of Electric Power Industry, Tokyo 201-8511, Japan}
\affiliation{Japan Atomic Energy Agency, Ibaraki 319-1195, Japan}

\begin{abstract}
A comprehensive investigation on point defects and their clustering behavior in nonstoichiometric
uranium dioxide UO$_{2\pm x}$ is carried out using LSDA+U method based on density functional
theory. Accurate energetic information and charge transfers available so far
are obtained. With these energies that have improved more than $50\%$ over
that of pure GGA and LDA, we show the density
functional theory predicts the predominance of oxygen defects over uranium ones at any compositions, which
is possible only after treated the localized 5\emph{f} electrons properly.
Calculations also suggest an upper
bound of $x\sim0.03$ for oxygen clusters to start off.
The volume change induced by point uranium defects is monotonic but nonlinear,
whereas for oxygen defects, increase $x$ always reduces the system volume linearly,
except dimers that require extra space for accommodation, which
has been identified as meta-stable ionic molecule.
Though oxygen dimers usually occupy Willis O$^{''}$ sites and mimic a single oxygen
in energetics and charge state, they are rare at ambient
conditions. Its decomposition process and vibrational properties
have been studied carefully.
To obtain a general clustering mechanism in anion-excess
fluorites systematically, we also analyze the local stabilities of possible basic
clustering modes of oxygen defects.
The result shows an unified way to understand the structure of Willis type
and cuboctahedral clusters in UO$_{2+x}$ and $\beta$-U$_{4}$O$_{9}$.
Finally we generalize the point defect model to the independent
clusters approximation to include clustering effects, the impact on defect populations is discussed.

\end{abstract}

\volumeyear{year}
\volumenumber{number}
\issuenumber{number}
\eid{identifier}
\maketitle


\section{INTRODUCTION}

Oxides of the fluorite structure include ZrO$_{2}$, a common ceramic in research
and industry, CeO$_{2}$, and the actinide oxides ThO$_{2}$, UO$_{2}$, and PuO$_{2}$.
The series of actinide dioxides is of great interest in nuclear applications. The
present generation of nuclear reactors uses UO$_{2}$ as nuclear fuel. Fast breeder
reactors at present employ mixed (U,\,Pu)O$_{2}$, and may in the future use
(U,\,Th)O$_{2}$. In the oxides of the fluorite or CaF$_{2}$ structure, MO$_{2}$,
each metal atom M is surrounded by eight equivalent nearest-neighbor O atoms
each of which in turn is surrounded by a tetrahedron of four equivalent M atoms.
A typical feature of the fluorite structure is the large ($\frac{1}{2}$,\,$\frac{1}{2}$,\,$\frac{1}{2}$)
octahedral holes in which interstitial ions can easily be accommodated.
Fluorite structure of UO$_{2}$
transforms to an orthorhombic \emph{Pnma} phase under a hydrostatic compression
beyond 40\,GPa, which in turn followed by an iso-structural
transition after 80\,GPa.\cite{idiri04,geng07}
At ambient pressure, however, it exists as the single phase, stoichiometric oxide at all
temperature up to 2073\,K. Above that it transforms to the sub-stoichiometric
phase UO$_{2-x}$, whereas at lower temperatures it easily dissolves large amounts
of interstitial oxygen to form \emph{anion-excess} compositions UO$_{2+x}$.
Higher interstitial concentration leads to another ordered phase U$_{4}$O$_{9}$,
which closely relates to fluorite structure.\cite{hering52}
It was argued that stoichiometric U$_{4}$O$_{9}$ does not exist, and should be
U$_{4}$O$_{9-y}$ actually.\cite{bevan86} But for simplicity we still
use U$_{4}$O$_{9}$ to refer to the non-stoichiometric phase hereinafter.
There are three polymorphs of U$_{4}$O$_{9}$ between room
temperature and 1273\,K, known as $\alpha$,\,$\beta$,\,$\gamma$,
with the $\alpha/\beta$ boundary at 353\,K and the $\beta/\gamma$ boundary at
about 873\,K. Only the detailed atomic arrangement in $\beta$-phase is
clearly determined: the excess anions accommodate in \emph{cuboctahedral clusters}
centered on the 12-fold sites of the cubic space group $I\overline{4}3d$ with
the uranium sublattice remains \emph{undisturbed}.\cite{cooper04,bevan86}
Although the unit cell is 64 times larger than a normal cubic fluorite cell, the \emph{average cell}
is still in fluorite-type except one has to introduce some vacancies at normal anions
sites and two types of interstitial oxygen, each sited about 1\,$\mathrm{\AA}$ from the
empty octahedral site of the FCC cation sublattice along $\langle110\rangle$ (O$^{'}$) and $\langle111\rangle$ (O$^{''}$) direction, respectively.
This characteristic is also shared by the $\alpha$-phase\cite{willis64a} and UO$_{2+x}$,\cite{willis64b,willis78}
with the difference that U$_{4}$O$_{9}$ has a long-range ordering for the interstitial
oxygen atoms while in UO$_{2+x}$ it is just short-range ordered.
To avoid some oxygens too close together, an intuitive proposal that different
kinds of oxygen defect are associated to form \emph{defect clusters} is widely
adopted when modeling these phases.\cite{willis64b,willis78}

At first sight the fact that interstitials were detected not at the body centers
of the cubic interstitial sites but at sites considerably displaced from this
symmetric position is puzzling. In rare earth doped alkaline earth fluorides
it has conclusively shown that at low interstitial
concentrations (1 mole \% or less) the anions occupy the symmetric body
center interstitial site, but usually the low-symmetry defect structure is a
general feature of anion-excess fluorites.\cite{catlow81} About half century has elapsed, people
still know few about the stabilization mechanism of Willis O$^{'}$ and O$^{''}$ sites
in energetics. In the limit of $x\rightarrow 0$ in UO$_{2+x}$, whether
the excess-anions will occupy the octahedral interstitial
site or not is still \emph{unclear}. On the other hand, though the occurrence
of cuboctahedral clusters in $\beta$-U$_{4}$O$_{9}$ has been confirmed
by experiments,
the geometry of defect clusters in low interstitial concentration regime is
unknown. One of the simplest model is to assume the Willis 2:2:2 cluster
(see Ref.[\onlinecite{willis78}] for its geometry)
can exist independently and distribute randomly in the material around this concentration. Allen
proposed a model for U$_{4}$O$_{9}$ in this line
by chaining 2:2:2 clusters along $\langle110\rangle$ direction.\cite{allen82}
Unfortunately his model is definitely wrong because the inconsistencies
with experimental facts in: ({\romannumeral 1}) leads to an exact stoichiometric
U$_{4}$O$_{9}$, which might not exist; ({\romannumeral 2}) no cuboctahedral
clusters can be formed in his arrangement; ({\romannumeral 3}) has an equal
concentration for O$^{'}$ and O$^{''}$ sites, against the measurements that O$^{''}$
position has a much lower occupancy.\cite{bevan86,cooper04}

Therefore to investigate the geometry and stability of possible defect clusters with a first
principles method is required, but it never be easy. The big unit cell of U$_{4}$O$_{9}$
and the shortage of information about atomic arrangement in UO$_{2+x}$ have restricted
most of attempts within \emph{point defect} approximation, and only formation energy of
simple intrinsic defects (Frenkel pairs and Schottky defect) were
calculated.\cite{petit98,crocombette01,freyss05} Applied these energies
to point defect model (PDM),\cite{matzke87,lidiard66} however, did not produce satisfactory
\emph{defect populations} --- uranium vacancy dominates in the hyperstoichiometric
regime, against the experimental anticipation.\cite{crocombette01,freyss05} The failure might be attributed to the limitation of the
PDM which assumes isolated non-interacting point defects, whereas in UO$_{2+x}$
this is impossible when $x\geq 0.03$, as we will show later.
Also it can arise from the inaccurate energies produced by the
local density approximation (LDA) or the generalized gradient approximation (GGA)
of the electronic density functional that has been proven failed to describe \emph{localized states}.\cite{geng07}
Nevertheless, some qualitative properties still can access by static calculations
within this model. For example the diffusion rate of interstitials
can be modeled simply by estimating the migration energy along all possible
paths that bridge the initial and finial interstitial positions,
which is readily computable by \emph{ab initio} nudged
elastic band (NEB) algorithm. For UO$_{2+x}$, the conclusion
is that a direct diffusion is almost prohibited and a normal oxygen on fluorite
lattice site must be involved as an intermediate process. That is, the
interstitial atom pushes a neighboring lattice oxygen into another
interstitial site and itself jumps into the vacancy thus created (\emph{interstitialcy} mechanism).\cite{durinck06}
The extreme of this process is, evidently, creating a transient oxygen dimer,
and thus sets up an \emph{upper bound} to the migration energy for thermodynamical diffusion of oxygens.
In order to keep the occurrence probability of oxygen dimer being consistent
with experimental observation in bulk U$_{4}$O$_{9}$,\cite{garrido06}
the energy required to form a such kind of dimer should be much larger
than the average migration energy.
But this has not yet been confirmed by
\emph{ab initio} calculations. Near to the surface of UO$_{2}$ that exposed to air, however,
oxygen dimer might become prevailing due to \emph{oxidations}. And their
stability in UO$_{2}$ matrix may shed some light on the mechanism of how the
material dissolve O$_{2}$
molecules into individual interstitials.
Also, it serves as to verify the Willis' assumption that \emph{each O$^{''}$ interstitial
has to be associating with one vacancy that occupies the nearest oxygen
site},\cite{willis64b} since otherwise they must form an oxygen dimer.

These motivate the research work of this paper that mainly focuses on:
({\romannumeral 1}) the stability of isolated point oxygen interstitial in UO$_{2+x}$ when
$x\rightarrow 0$; ({\romannumeral 2}) the stability and decomposition process of oxygen dimer,
including the variations of energy, cell volume and charges, respectively; ({\romannumeral 3})
the local stability of defect clusters that composed of oxygen vacancies, O$^{'}$ and O$^{''}$
interstitials. These clusters can be viewed as fractal pieces of a cuboctahedral
cluster, which is the essential in U$_{4}$O$_{9}$ phase. It is believed that the
transition from UO$_{2+x}$ to U$_{4}$O$_{9}$ involves \emph{long-range ordering} of
the defect complexes, leading to a change in the symmetry relating the relative
positions of the complexes, without producing any atomic re-arrangement within
these complexes, \emph{i.e}., micro-domains of U$_{4}$O$_{9}$ should already exist in
UO$_{2+x}$.\cite{willis64b} What we also want to find out primarily in this paper is
what kind of cluster is the most possible candidate for
this complex, and its polymorphs when $x$ is increased.
In next section we will brief the calculation method.
Main results and discussions are presented in section \ref{sec:discuss}:
Sec.\ref{sec:formE} devotes to formation energy analysis and Sec.\ref{sec:charge}
the charge transfers, in Sec.\ref{sec:O-dimer} and \ref{sec:vibf}
we will discuss the properties of oxygen dimer in UO$_{2}$ and its decomposition
process. The defect clustering pattern and its tendency with increased $x$ are given
in Sec.\ref{sec:clustering}, while in Sec.\ref{sec:concentration} a generalization
of PDM to include clustering effects is proposed, as well as the associated defect population analysis. Finally in
Sec.\ref{sec:conclu} we summarize the paper.

\section{METHOD OF CALCULATION}

Our investigation on defective behavior of UO$_{2}$ based on a series of total energy calculations
with different configurations in fluorite
structure which varied in simulation cell size and defect arrangement. The plane-wave method
using density functional theory (DFT) to treat the electronic energy as implemented
in VASP code\cite{vasp,kresse96} was employed, as well as the projector-augmented
wave (PAW) pseudopotentials.\cite{blochl94,kresse99}
The $2s^{2}2p^{4}$ electrons in oxygen and $6s^{2}6p^{6}5f^{3}6d^{1}7s^{2}$
in uranium were treated in valence space.
The cutoff for kinetic energy of plane waves was set as high as 500\,eV to eliminate
the possible Pulay stress erroneous.
Also it has been elevated due to the presence of oxygen which requires an energy cutoff at least
400\,eV to converge the electronic energy within a few $m$eV.
Integrations over reciprocal space were performed in the irreducible
Brillouin zone with about 8$\sim$36 non-equivalent k-points,
depending on the system size.
The energy tolerance for
charge self-consistency convergence was set to $1\times 10^{-5}$\,eV for all
calculations.
And the total convergence of this parameter set was checked well.
Without a specific statement, all structures in following discussions have been fully
relaxed to get all Hellman-Feynman forces (stress) less than $0.01$\,eV/\AA.

The electronic exchange-correlation energy was computed by spin-polarized local
density approximation with an
effective on-site Coulomb interaction to split the partially filled 5\emph{f}
bands localized on uranium atoms (LSDA+U).\cite{anisimov91,anisimov93}
Parameters of the Hubbard
term were taken as $U=4.5$\,eV and $J=0.51$\,eV, which has been checked carefully
for \emph{fluorite} UO$_{2}$.\cite{dudarev00,dudarev97,dudarev98,geng07}
Here some comments are desired. It is well known that it is the $U$ but
$J$ that contributes to electronic structure sensitively.
In UO$_{2}$ case, the value of $U$ quite depends on the atomic arrangement of uranium atoms.\cite{geng07}
If uranium sublattice almost being unchanged, as the case
here concerned, one can expect the $U$ would not vary too much. On the other hand,
the influence of interstitial
oxygens on localized 5$f$ electrons should be small if they are well separated
from uranium atoms.
However, as interstitial concentration increased, the impact on $U$ may become
non-negligible. Therefore we must restrict to certain composition regime, and
$x\leq 0.25$ should be small enough to allow using this set of parameters.
This composition value can be estimated roughly by checking the induced deformation
on the uranium sublattice.
The situation of uranium defects is a little embarrassed. We cannot estimate
its effect on $U$ until a more accurate functional becomes generally available,
for example, the hybrid density functional that has shown impressive versatility
in preliminary applications.\cite{hybrid}
However, for a point defect in a large enough cell, to neglect this influence seems reasonable.
Another point is about the adoption of LSDA+U functional instead of GGA+U. The latter
has been proven as of a poor description to the defect energetics, which we will discuss
in details in Sec.\ref{sec:concentration}.

\begin{table}[]
\caption{\label{tab:coheE} Equilibrium properties of uranium dioxide with defects:
superscript $u$ denotes uranium defects and negative subscript refers to
vacancy.
$\Delta V$ is the volume difference relative to $C1$ structure and $E_{f}$ the defect
formation energy per point defect. Note $\overline{E}_{coh}$ and volume have been averaged
to a single fluorite cubic cell.}
\begin{ruledtabular}
\begin{tabular}{l c c c c l} 
  Label & $\overline{E}_{coh}$(eV/cell) & Volume(\AA$^{3}$/cell) & $\Delta V$(\AA$^{3}$/cell) & E$_{f}$(eV) & Structure \\
\hline
  $C1$ & -98.638 & 161.34 & 0.0 & 0.0 & \includegraphics*[scale=0.5]{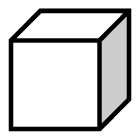} \\
  $C1_{1}$ & -102.906     & 157.17 & -4.17 & -1.394  & \includegraphics*[scale=0.5]{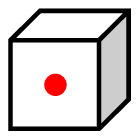} \\
  $C2_{1}$(L/S) & -101.20/-101.199   & 159.47/159.46 &-1.87/-1.88  &-2.249/-2.248    & \includegraphics*[scale=0.5]{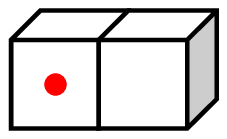} \\
  $C4_{1}$(L/S) & -99.71/-99.731   & 160.54/160.28 &-0.8/-1.06  &-1.413/-1.496  & \includegraphics*[scale=0.5]{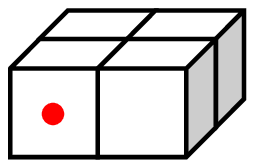}\\
  $C4_{1d}$(S)& -99.337  & 162.87 & 1.53 & 0.079 & \includegraphics*[scale=0.5]{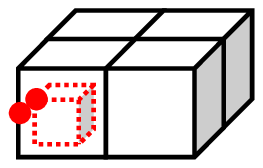}\\
  $C4_{2d}$(S)& -100.486   & 163.05 &1.71  & -1.642\footnotemark[1]{} & \includegraphics*[scale=0.5]{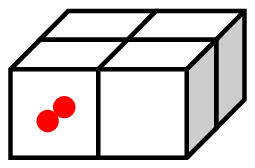} \\
  $C4_{1d1}$(S)& -100.461  & 162.09 & 0.75 & -1.545\footnotemark[1]{} & \includegraphics*[scale=0.5]{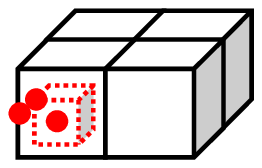} \\
  $C4_{2}$(L/S)& -101.233/-101.237  & 159.35/159.38 & -1.99/-1.96 & -2.316/-2.324 & \includegraphics*[scale=0.5]{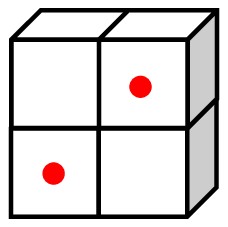} \\
  $C8_{1}$& -99.268    & 161.05 & -0.29 & -2.169 & \includegraphics*[scale=0.5]{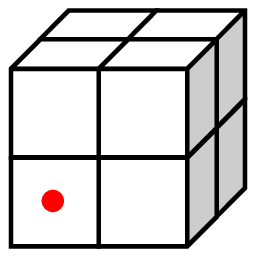} \\
  $C3_{1}^{L}$(L/S)& -100.099/-100.361 & 160.25/160.16 & -1.09/-1.18  &-1.509/-2.294  & \includegraphics*[scale=0.5]{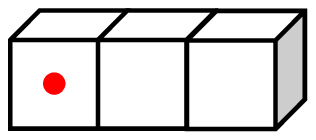} \\
  $C3_{2}^{L}$(L/S)& -101.789/-101.788 & 159.34/159.36 & -2.0/-1.98 & -1.853/-1.850 & \includegraphics*[scale=0.5]{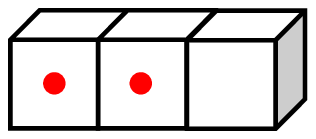} \\
  $C8_{-1}$&-97.338 & 161.54 & 0.20 &  7.525 & $-$  \\
  ${}^{u}C8_{1}$&-98.289 & 164.25 & 2.91 &8.194  & $-$ \\
  ${}^{u}C8_{-1}$&-96.831 & 160.26 & -1.08 &  9.056& $-$ \\
\end{tabular}
\footnotetext[1]{\,per two oxygen interstitials}\\
\end{ruledtabular}
\end{table}

The \emph{supercell} method has been used to model defect structures.
Periodic boundary conditions were imposed on the whole system.
The geometry of all structures (except those in Sec.\ref{sec:clustering}) are listed in table \ref{tab:coheE}, where each brick
indicates a fluorite cubic unit cell (in U$_{4}$O$_{8}$), and red points represent oxygen interstitials,
which usually occupy the cubic centers, except those associated with dimers.
Dot-lined box (if drawn) indicates the oxygen cage.
No atom on the fluorite lattice has been drawn explicitly, except in $C4_{1d}$ and
$C4_{1d1}$ where the lattice oxygens bonded to interstitials were also plotted.
Each structure of $C4_{1d}$,
$C4_{2d}$ and $C4_{1d1}$ contains one oxygen dimer, respectively.
Configuration $^{u}C8_{1}$ has the same geometry as $C8_{1}$ but replacing
the interstitial oxygen with one uranium, and $^{u}C8_{-1}$ or $C8_{-1}$
corresponds to remove one lattice atom from a system with 8 fluorite cubic cells ($2\times2\times2$).

The magnetic effects have been taken into account by initially set up an
\emph{antiferromagnetic} orientation of atomic moments. Two cases, the
moment ordering along the longest (L) and shortest (S) axis, are considered.
The cohesive energy $E_{coh}$ of each structure is calculated from the total energy
by subtracting the isolated spin-polarized atomic contributions. Then
the oxygen defect formation energy in structure $Cm_{n}$ is given by
\begin{equation}
  E_{f}=E_{coh}-mE_{coh}^{C1}-\frac{n}{2}E_{O_{2}}.
\end{equation}
Here $m$ is the number of fluorite cubic cells and $n$ the total oxygen interstitials or
vacancies.
$E_{O_{2}}$ is the binding energy of a neutral dioxygen molecule. Alternatively,
one can define an \emph{alloy-system} like formation energy by
choosing $C1_{1}$ as one of the reference phases instead of O$_{2}$ molecule.
We call it the \emph{relative formation energy}, which
takes the advantage of showing the phase stability of superstructures
with different composition explicitly, analogous to that in an alloy and compound system.\cite{geng04,geng05}
It thus can be calculated as
\begin{equation}
  E_{Rf}=E_{coh}-(1-\frac{n}{m})E_{coh}^{C1}-\frac{n}{m}E_{coh}^{C1_{1}},
  \label{eq:rf}
\end{equation}
and the value of ${n}/{m}$ stands for the composition of phase $C1_{1}$
in $C1$, or equivalently, the concentration of oxygen interstitial per
fluorite cubic cell.
All configurations incorporated with uranium defect are marked by a superscript $u$ in table \ref{tab:coheE}, and the formation
energy for a defect in $^{u}Cm_{n}$ is defined as
\begin{equation}
E_{f}=E_{coh}-mE_{coh}^{C1}-nE_{\alpha U}.
\end{equation}
Here $E_{\alpha U}$ is the cohesive energy per atom in the metallic $\alpha$\emph{-U} phase,
and we use the experimental value of $-5.4$\,eV for simplicity.\cite{kittel96}

Vibrational frequencies of interstitial oxygens were calculated by finite
difference method with \emph{frozen phonon} approximation.
At finite temperatures, these vibrational frequencies contribute to the first order of
defect free energy directly, which is given by
$F(T)=E_{f}-\kappa_{B} T\ln Z_{v}$, with the partition function
\begin{equation}
Z_{v}=\prod_{i}\sum_{j=0}^{\infty}\exp\left(\frac{-E_{j}^{i}}{\kappa_{B} T}\right),
\label{eq:freeE}
\end{equation}
where $\kappa_{B}$ is the Boltzmann constant and $E_{j}^{i}$ the eigenvalue energy for the $j$-th vibrational mode with
frequency $\omega^{i}$,
and the harmonic approximation $E_{j}^{i}=\hbar \omega^{i}(j+\frac{1}{2})$ has been used.
Mind here we have not subtracted the vibrational free energy of the reference state
O$_{2}$ molecule, and comparison of the calculated free energies therefore can be made only among configurations
with the \emph{same} number of interstitial oxygens.

Regarding charge transfer calculations, it is well known that
the concept of static atomic charge in \emph{ab initio} calculations usually leads to
ambiguity due to the arbitrariness in determining the belongingness of electrons.
Nevertheless, there are several methods exist to compute the effective atomic charge,
which do provide some useful qualitative understanding.
Among those the Bader's conception that to partition an electronic density
by surfaces formed by the density minimums (zero flux surfaces)
is one of the most intuitive. It is simple to calculate Bader charges,
requiring only atomic positions and electronic density as input. The
partition surfaces are determined by finding the charge density minimums.\cite{henkel06}
Then the atomic
charge is obtained by subtracting the valent electrons from the integral of
charge density over the space surrounded by the partition surfaces
that envelops the atom.
Another widely used concept is dynamical effective charge, defined by the change
of polarization induced by atomic displacements,\cite{ghosez98} which is beyond the scope of this
paper and will not elaborate here.

\section{RESULTS AND DISCUSSIONS}
\label{sec:discuss}
\subsection{Dioxygen molecule}

We first discuss the dioxygen molecule. The O$_{2}$ molecule was modeled by putting it in a periodic cubic cell with a lattice constant
of 15\,\AA, large enough to eliminate the factitious interaction among its images. Only
one k point ($\Gamma$) was used. Since the notorious failure of LDA in description small isolated molecules,
we employed here (and only here) the revised Perdew-Burke-Ernzerhof (rPBE)\cite{hammer99} GGA electronic
exchange-correlation
functional. The bond length was optimized to
be 1.22\,\AA, in a good agreement with experimental 1.21\,\AA.\cite{huber79,Jolly85}
The calculated binding energy is $-5.75$\,eV, a little deeper than observed
$-5.1$\,eV.\cite{herzberg52} This discrepancy should attribute to the difficulty of
current functional to take into account the van der Waals interactions accurately. The vibrational
frequency of stretch mode, however, was well reproduced as 1588.6\,cm$^{-1}$,
against the experimental 1580.2\,cm$^{-1}$.\cite{huber79}
As a check to the validity of Bader's conception, we calculated the Bader
atomic charge for each oxygen atom in O$_{2}$ and got them as $\pm 0.09\,e$, reflects the essential of \emph{covalent} bond correctly.
The deviation can be reduced further when in an ionic bond environment where the
charge density minimum surfaces sharply show up.

\subsection{Structure and formation energies}
\label{sec:formE}
\subsubsection{Oxygen interstitials}

The calculated equilibrium properties of 14 configurations, including cohesive
energies, equilibrium volumes, volume changes relative to the ideal UO$_{2}$ cell and
defect formation energies are listed in table \ref{tab:coheE}. These data have been \emph{averaged}
to one fluorite cubic cell. It can be seen that the cohesive energy always
decreases as oxygen interstitial concentration increased, demonstrating the
tendency of uranium dioxide to dissolve oxygens. The solubility, however, cannot
be determined by simply taking the limit of this cohesive energy \emph{vs} concentration curve.
Also, the relative stability
among different configurations has been obscured here. To get that information explicitly,
one needs back to the relative formation energy.

\begin{figure}
  \includegraphics*[0.22in,0.19in][3.9in,2.96in]{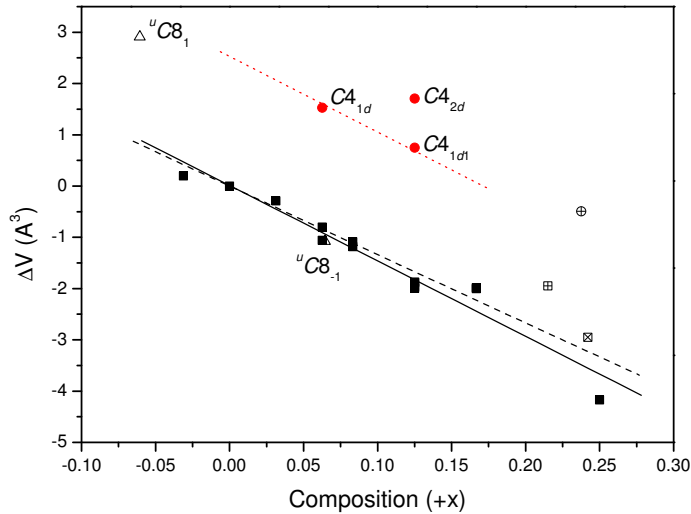}
  \caption{(Color online) Calculated variation of the volume change in UO$_{2+x}$
  with the deviation from stoichiometry $x$: Solid squares stand for point oxygen defects,
  and the solid line is the linear fitting to them; open triangles are for uranium
  defects; solid circles are those incorporated with one oxygen dimer; the dotted line
  is for eye guide. Experimental data: dashed line is for UO$_{2+x}$ reported
  by Alekseyev \emph{et al.} and others for $\beta$-U$_{4}$O$_{9}$ (at
  room temperature $\boxtimes$,
  503\,K $\boxplus$, and 773\,K $\oplus$, respectively.)}
  \label{fig:volume}
\end{figure}

One interesting thing is that we find except that of oxygen dimer, introducing
point oxygen interstitials always shrinks the system, \emph{i.e.}, leads to a \emph{negative}
$\Delta V$, as shown in figure \ref{fig:volume}. This feature differs from GGA results,\cite{freyss05} but agrees with
GGA+U,\cite{iwasawa06} and may attribute to the behavior of localized 5$f$ electrons.
Generally, a negative $\Delta V$ means the interaction between the matrix
and the interstitials is dominated by attractive chemical potentials rather than by mechanical
effects (atomic \emph{size effect}).
The latter always results in a swollen volume and is important for big
interstitial atoms or inert gases. Oxygen dimer belongs to this class and
requires extra space to accommodate, which can be seen
more clearly when compare $C4_{1d}$ with $C4_{1}$ and $C4_{1d1}$ configurations.
The influence of magnetic orientation on equilibrium volume is almost negligible except in
the cases of $C4_{1}$ and $C3_{1}^{L}$, of which only $C3_{1}^{L}$ has a notable
formation energy difference between L and S orientation.

The calculated slope of volume variation induced by oxygen interstitials (the solid
line in figure \ref{fig:volume}) is in a good agreement with experimental change
of the lattice constant $a=5.4696-0.1495x$
as reported by Alekseyev \emph{et al.} for homogenous UO$_{2+x}$ powders as
quoted in Ref.[\onlinecite{McEachern98}] (the dashed line).
Also, it is in accord with the volume change of $\beta$-U$_{4}$O$_{9}$
measured at room temperature\cite{garrido06} with respect to that of
stoichiometric UO$_{2}.$\cite{idiri04} Increase temperature to 503\,K and
773\,K expands the material greatly,\cite{cooper04} which can be understood in terms of
thermal vibration effects and extensive defects generation.

\begin{figure}
  \includegraphics*[0.22in,0.19in][3.9in,2.96in]{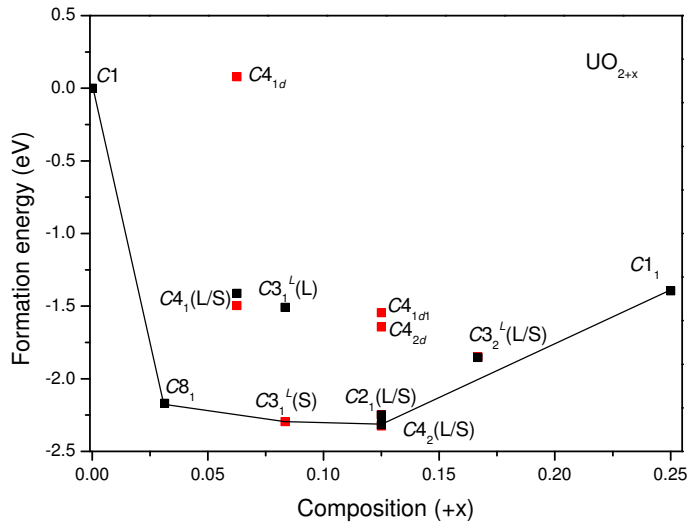}
  \caption{(Color online) Formation energy of oxygen interstitials in UO$_{2}$ arranged in various
  configurations. $C8_{1}$ corresponds to an isolated defect approximation and other configurations
  must be interpreted as ordered defect phases.}
  \label{fig:form-E}
\end{figure}

Figure \ref{fig:form-E} shows the defect formation energies of oxygen interstitial in all considered
configurations of UO$_{2+x}$ within $0\leq x\leq 0.25$. Note the value of $C4_{2d}$
and $C4_{1d1}$ are for two interstitials. A remarkable feature in this graph
is that in \emph{energetics} an oxygen dimer mimics a single oxygen atom. Comparing
that of the perfect crystal $C1$ with $C4_{1d}$, and that of $C4_{1}$ with $C4_{2d}$ and $C4_{1d1}$,
we see that despite the latter contains one more interstitial, the formation energy
is almost the same. That means to absorb an oxygen from O$_{2}$ gas into UO$_{2}$ and
forms a dimer will neither release nor gain heat. Point interstitial and dimer would have almost the
same behavior except that a dimer needs a bigger space for accommodation. This mimic is
also supported by Bader effective charge calculations: they have almost the same charge too (see below).
However, this does not suggest the stability of oxygen dimers in UO$_{2}$ since
point oxygen interstitial always has a lower per atom formation energy.

Our calculations also present a remarkable system size dependence in formation energy,
contrasts to that of GGA results where a value of $-2.6$ and $-2.5$\,eV were
obtained for $C1_{1}$ and $C2_{1}$ configurations (almost size-independent), respectively,\cite{freyss05}
revealing the limitation of applying the pure GGA to defects in spite of its impressive
performance in energetics of perfect bulk UO$_{2}$.\cite{geng07} No magnetic
ordering and volume relaxation were considered in that GGA calculation.\cite{freyss05} A
discrepancy about 1.2\,eV
with our result for $C1_{1}$, however, cannot be covered by these
effects since volume relaxation would definitely increase the discrepancy
and magnetic contribution can not be of that magnitude,
and it therefore should attribute to the behavior of localized 5\emph{f} states.

\begin{figure}
  \includegraphics*[0.22in,0.19in][3.9in,2.96in]{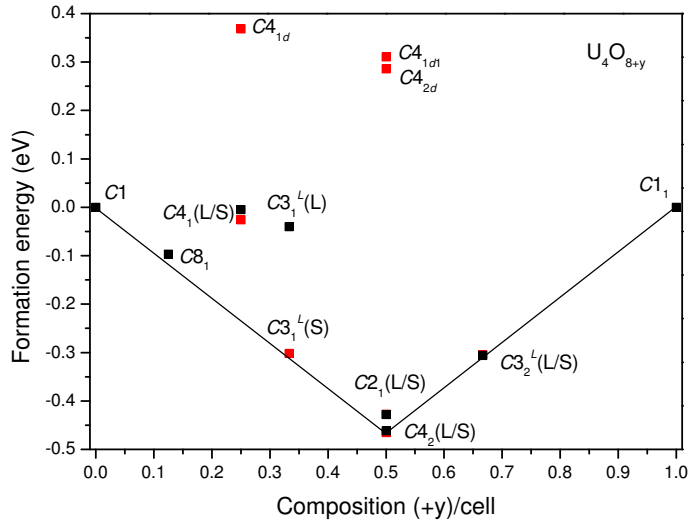}
  \caption{(Color online) Relative formation energy of different phases in U$_{4}$O$_{8+y}$.
   $C8_{1}$ corresponds to an isolated defect approximation and other configurations
  must be interpreted as ordered defect phases.}
  \label{fig:rform-E}
\end{figure}

The deepest formation energy shown in figure \ref{fig:form-E} is $-2.32$\,eV (configuration $C4_{2}$), rather than
the isolated approximation of a point interstitial's ($C8_{1}$) $-2.17$\,eV.
Actually, except those configurations with 8 fluorite cubic cells, the
defects in all other structures cannot be interpreted as isolated ones because
the non-negligible interactions among their images arisen from periodic conditions.
This invalidates the defect stability analysis based on their
formation energy directly.
Mapping these configurations onto an alloy system can circumvent this difficulty.
Namely, to view these configurations (discard those
with dimer) as an alloy system with oxygen interstitials distributing
over the fluorite cubic centers (U$_{4}$O$_{8+y}$). Then the extreme phases of this system are $C1$ and $C1_{1}$.
Following this way, figure \ref{fig:form-E}
transforms into figure \ref{fig:rform-E} with the help of Eq.(\ref{eq:rf}), where the sold line indicates the ground
state hull. We then find that $C3_{1}^{L}(\mathrm{S})$ and $C3_{2}^{L}$ are
close to be ground states, while $C8_{1}$, the isolated point interstitial approximation,
will decompose into a mixture of $C1$ and $C4_{2}$ phases. That means, defect clustering
is \emph{inevitable} when $x\geq 0.03$.\cite{note0} Since $C4_{2}$ may not be the physical ground state
(mind neutron diffraction experiments suggested that no octahedral site
should be occupied around this composition\cite{willis64b,willis78}), this limit can be
lowered further.
On the other hand, it seems reasonable to assume that $C8_{1}$ has already approached
the limit of an isolated point interstitial, namely, no notable formation energy
would be gained or lost if enlarge the system to $C27_{1}$ or $C64_{1}$. If it is true,
then the isolated point interstitial will always exist when $x\rightarrow 0$. Its site, according
to structure symmetry, should be the octahedral position.
It is worthy to note that the PDM fails at about $x\sim 10^{-4}$ instead of $10^{-2}$ with
GGA formation energies.\cite{freyss05} This two orders discrepancy is owing to the inaccuracy
of the formation energies they used, which can be improved greatly by LSDA+U method,
see Sec.\ref{sec:concentration} for details.

\subsubsection{Other defects}

Point oxygen vacancy and uranium defects are all modeled in a system with 8 fluorite
cubic cells. Namely by $C8_{-1}$ and $^{u}C8_{\pm1}$.
The volume change in $C8_{-1}$ is in accord with that of point oxygen interstitials:
linearly decreasing with an increased $x$, fitting to $\triangle V=0.01-14.7x$,
as shown by the solid line in figure \ref{fig:volume}. Uranium vacancy also obeys
this law, whereas the interstitial has a much rapid change. Totally, they
still decrease monotonically with $x$, but no longer in linear. All three
defects have a formation energy larger than 7\,eV, in contrast to previous
\emph{ab initio} results. We prefer to postpone this discussion to Sec.\ref{sec:concentration}
with Frenkel and Schottky defects together.

\subsection{Charge transfers}
\label{sec:charge}

\begin{table}[]
\caption{\label{tab:charge} Bader effective charges of UO$_{2}$ with defects:
average charge $q$, standard deviation $\sigma$, difference from that in perfect UO$_{2}$ $\delta q$
and the maximal transferred charge $\Delta_{max}$ ($\pm0.02$). All in the unit of positron charge $e$.}
\begin{ruledtabular}
\begin{tabular}{l c c c c c c c c c} 
  Label & defects & \multicolumn{4}{c}{uranium} & \multicolumn{4}{c}{oxygen} \\
  \cline{3-6}\cline{7-10}
     & $q$  &$q$ & $\sigma$&$\delta q$ & $\Delta_{max}$ &$q$ & $\sigma$&$\delta q$ & $\Delta_{max}$\\
\hline
  $C1$         & $-$ &2.56 & 0.0 & 0.0 & 0.0 & -1.28  & 0.0 & 0.0 & 0.0 \\
  $C1_{1}$     & -1.04  & 2.62& 0.11 &  0.07 & 0.26 & -1.18  & 0.004& 0.10 & 0.11\\
  $C2_{1}$(L/S) & -1.15   & 2.63 & 0.11 & 0.08   &0.27 & -1.24 & 0.02 & 0.03 & 0.07  \\
  $C4_{1}$(L/S) &  -1.18/-1.14  & 2.60 & 0.08 & 0.04 & 0.25 & -1.26 & 0.03 & 0.02 & 0.08 \\
  $C4_{1d}$(S)& -0.61(-0.77)\footnotemark[1]{} & 2.56 & 0.02 & 0.01 & -0.03 & -1.28 & 0.01 & 0.0 & -0.03 \\
  $C4_{2d}$(S)& -0.66,\,-0.59   & 2.60 & 0.08 & 0.04 & 0.26 & -1.26 & 0.03 & 0.02 & 0.12 \\
  $C4_{2}$(L/S)&  -1.19/-1.20   &2.61/2.63  &0.10  &0.07  &0.23/0.25 & -1.24 &0.02 &0.04 &0.07  \\
  $C8_{1}$&  -1.24   & 2.58 & 0.05 & 0.03 & 0.24 & -1.27 & 0.02 & 0.01 & 0.05 \\
  $C3_{1}^{L}$(L/S)& -1.16 & 2.60 & 0.09  & 0.04 &0.25 & -1.25 & 0.03 & 0.03 & 0.10/0.08 \\
  $C3_{2}^{L}$(L/S)& -1.10,\,-1.13 & 2.64 & 0.11 & 0.09 & 0.26/0.28 & -1.23 & 0.02 & 0.05 & 0.09 \\
 $C8_{-1}$& $-$ &2.53  & 0.09 & -0.03 & -0.34 & -1.28 & 0.01 & -0.00& -0.03\\
 ${}^{u}C8_{1}$&1.61  &2.51  &0.09  &-0.04  &-0.25&-1.28&0.01&-0.00&-0.03 \\
 ${}^{u}C8_{-1}$& $-$ & 2.59 & 0.08 & 0.04 &0.26 &-1.26& 0.03& 0.02 &0.13 \\
\end{tabular}
\footnotetext[1]{\,value in parenthesis is for the atom sited on oxygen
sublattice}\\
\end{ruledtabular}
\end{table}

It has long been believed that dissolve oxygen in UO$_{2}$ will \emph{oxidize} U$^{4+}$
to U$^{5+}$, even U$^{6+}$ state. The exact charge transfer induced by defects, however,
is unclear. Qualitative analysis is accessible to this problem with empirical
shell model,\cite{catlow77} nevertheless the calculated energy depends on atomic positions
sensitively,\cite{geng07b} obscuring its applicability to defects with
noticeable structure deformations. A direct calculation of the
charge state from first principles is therefore desired.

\subsubsection{Oxygen interstitials}

The calculated Bader effective charges using electronic density generated
with VASP code are listed in table \ref{tab:charge}, where the interstitials
and the lattice oxygens that forming a dimer are excluded from the average operations, and listed
separately in the \emph{defects} column.
We find all oxygen interstitials that occupy the cubic body center
having a charge state close to the lattice oxygens, especially in the $C8_{1}$
phase where the difference is only 0.03\,$e$. In $C8_{1}$ the disturbance
to lattice oxygens is also small, the largest charge transfer
is just 0.05\,$e$. A similar situation holds for uranium atoms, except
two of them lost about 0.24\,$e$, respectively, which contribute to
the standard deviation directly. Considering oxygen and uranium
in perfect UO$_{2}$ have only a charge of $-1.28$ and $2.56$\,$e$, all are smaller
than the nominal chemical valences but close to that of a partially ionic model
that widely used in semi-empirical potentials,\cite{geng07b,yamada00} we
can re-interpret the Bader charges by multiplying a scaling factor
to make them comparable with the chemical valences. In this sense
the change of the charge state in these two uraniums should be about 0.5\,$e$, \emph{i.e}.,
they are oxidized to U$^{4.5+}$ instead of U$^{5+}$. The transferred charges,
however, cannot cover the amount absorbed by the interstitial oxygen, and all other
normal uraniums and oxygens have also lost a small portion of their charge.
This observation contrasts to the conventional expectation and reveals the difficulty
to oxidize uranium to a higher valence state. The charge transfers in other
configurations also support this point: in all cases each oxygen interstitial
can oxidize \emph{two} and only two uraniums to U$^{4.5+}$ while leaving others almost unchanged,
no higher valence state of uranium has been observed.
As to which uranium
is apt to be oxidized, obviously the answer is the nearest neighbors (NN)
of the defect, but oxidization of some next NNs also was
observed. It is worthwhile to point out that we did not find a sensitive dependence
of the charge state on Hubbard \emph{U} parameter.

The more deformed the geometry is, or equivalently, the more interstitials the system contains,
the charge state of lattice atoms are disturbed more drastically.
It is clear by comparing the charge transfers in $C3_{1}^{L}$ with
$C3_{2}^{L}$, or $C4_{1}$ with $C4_{2}$. The
largest $\Delta_{max}$ for oxygen takes place in $C1_{1}$ with the
largest composition, and in $C4_{2d}$ with a dimer. The smallest $\Delta_{max}$
for uranium and oxygen are in $C4_{1d}$,  also containing
a dimer, both are $-0.03\,e$. The difference between $C4_{1d}$ and
$C4_{2d}$ is the former contains only one interstitial which bonds to a
lattice oxygen and the latter contains two interstitials that bonding to
each other. Table \ref{tab:charge} illustrates that in the former case no charge
has been transferred from other lattice atoms, and only charge redistribution within
the dimer is involved that making it has a total charge close to a lattice oxygen;
in the latter case, however, absorbing charges from other atoms is necessary
and gives them a similar charge state as the interstitial in $C4_{1}$, especially only
two uraniums are oxidized to U$^{4.5+}$ state here in spite of there are two
interstitials presented. The total charge of the dimer, $-1.25\,e$, close
to a lattice oxygen in UO$_{2}$, indicates it should be O$_{2}^{2-}$ actually.

It is worthwhile to note that oxygen changes its charge state almost continuously
but it is \emph{discrete} for uranium when lost its charge. That is, except those atoms who lost $\sim0.25\,e$,
the changes of charge in other uraniums are less than $0.03\,e$. Moreover
the discrete lose of charge is always accompanied by lowering the local moment of uranium from
$\sim 2\,\mu_{\mathrm{B}}$ to $\sim 1\,\mu_{\mathrm{B}}$. Since the local moment
of uranium in UO$_{2}$ originates from localized 5\emph{f} states, it is
obvious that 5$f$ electrons contribute to this process greatly. This can be
understood in the partially ionic charge model: although the chemical valence
of uranium in UO$_{2}$ is $4+$, table \ref{tab:charge} shows in fact the
physical valence has only $2.56+$. Namely, only the $7s^{2}$ and
a fraction of $6d^{1}$ electrons are
completely transferred to oxygen. Uranium cation still holds about $0.24\,e$
of the $6d^{1}$ electron and other remainder forms two weak U-O
covalent bonds, each has a portion of $\sim 0.2\,e$. When oxidized by oxygen
interstitials, the cation loses its $6d^{1}$ electron completely (transferred
to the interstitial atoms). As a consequence one of the localized 5\emph{f} states becomes
the outermost orbit, which spreads extensively and eventually the cation
lost half of its local moment. This mechanism also explains the difficulty
to oxidize uranium to a higher charge state since transferring a 5\emph{f}
electron requires much larger energy than 6\emph{d} one.

\subsubsection{Other defects}

In point oxygen vacancy case (configuration $C8_{-1}$), the uranium cations gain
charges and decrease the average valence to $2.53+$, but the disturbance to remain
oxygen is small. The largest charge transfer for uranium is $-0.34\,e$, associating
with three other uranium atoms each of them gets an extra charge about $-0.25\,e$, respectively. Compared
with the interstitial case, here no notable change in local moment was
observed. The value of $-0.34\,e$ implies the cation has retracted the portion
of electrons shared by the removed oxygen ($\sim0.1\,e$), and the $-0.25\,e$ indicates
that each quarter-filled states of $6d$ electrons seems stabler than continuous occupancy.

Point uranium vacancy is analogous to two oxygen interstitials in that there are
four uranium cations lost the charge, three NNs and one next next NN, ranging from 0.23 to 0.26\,$e$.
All of them also lost half of their local moments. The change in other uraniums is negligible.
However, it disturbs the oxygens
severely, with a $\Delta_{max}$ as high as 0.13\,$e$, even though the averaged
charge is still close to the perfect one. The oxygen charge state in $^{u}C8_{1}$
is almost the same as in $C8_{-1}$, except that here there are six (NNs) instead four
uraniums gain charge, ranging from $-0.19\sim-0.25$\,$e$. Again, no apparent
impact on other atoms. The extra charge provided by the interstitial uranium
is almost absorbed by its six NNs completely. The magnetic ordering
has been damaged severely, and the change in exchange interaction has made some 5\emph{f} electrons
flip their spins, but no uranium was observed to have a moment of $\sim 1\,\mu_{\mathrm{B}}$.

\subsection{Oxygen dimer in UO$_{2}$}
\label{sec:O-dimer}

\begin{figure}
  \includegraphics*[0.22in,0.19in][4.34in,2.96in]{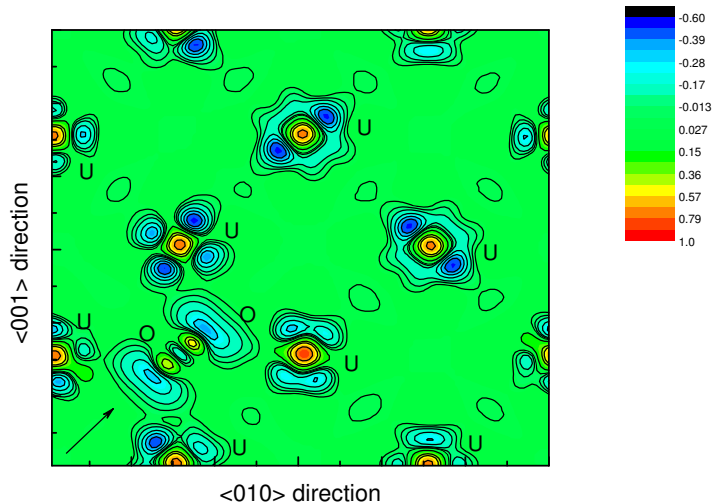}
  \caption{(Color online) The difference charge density of an oxygen dimer
  in $C4_{2d}$ configuration projected onto the [100] plane crossing the dimer center.
  An analogous density holds for the dimer in $C4_{1d}$.
}
  \label{fig:dimer-chgdif}
\end{figure}

As previous sections mentioned, although oxygen dimer has a similar
behavior in energetics and charge state as a single oxygen interstitial,
it actually is an ionic molecule,
and formed when oxygens are forced to close to each other enough.
But this is difficult due to the energy barrier between the individual atoms.
In UO$_{2}$, irradiation provides enough excess
energy to overcome this barrier. For example
in an $\alpha$ decay the recoil of the daughter nucleus produces a ballistic
shock with an energy release of about 70 keV,\cite{weber84,robinson94}
which frequently takes place in nuclear fuels.
Nonetheless this cannot survive the dimers to equilibrium conditions, even if
they do appear transiently.
Another situation where oxygen dimers
can be observed is near the \emph{surfaces} exposed to oxygen gas. Oxygen molecules
adsorbed onto the UO$_{2}$ surface will obtain additional charges then diffuse inwards.
Decomposing the molecule at the vacuum side of the surface is almost impossible due to the
large binding energy, while in UO$_{2}$ side it prefers to oxygen
sublattice sites instead of the interstitial positions,
where it decomposes into individual interstitials, with a barrier only
about $0.21$\,eV (see below).\cite{note1}

Figure \ref{fig:dimer-chgdif} shows the difference charge density (reference to the corresponding atomic charge) of an oxygen dimer
in UO$_{2}$ ($C4_{2d}$) projected to [100] plane, as the arrow indicates. The
covalent bond between two interstitial oxygens presents evidently. A similar
picture has been observed in $C4_{1d}$ configuration or a natural O$_{2}$ molecule.
Analysis shows that it
in fact is an O$_{2}^{2-}$, with the two additional electrons occupying the $2p\pi^{*}$
antibonding orbitals and the final bond order is one. The calculated bond length
is $1.39$\AA, slightly short than experimental $1.49$\AA.\cite{Jolly85}
This discrepancy owes to the compression from oxygen cage and
can be removed.
For example when the dimer is formed by bonding to one lattice oxygen ($C4_{1d}$),
where although the charge state is still similar ($-1.38$\,$e$), the bond length
extends to $1.47$\,\AA, in a good agreement with experimental data.

\begin{figure}
  \includegraphics*[0.22in,0.19in][4.38in,2.96in]{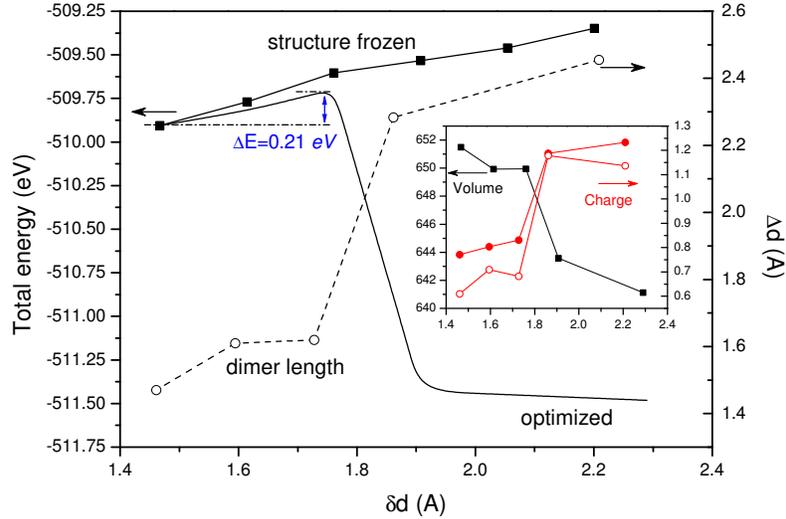}
  \caption{(Color online) Behavior of oxygens during a dimer decomposing process
  along $\langle111\rangle$:
  potential shape (solid lines) and equilibrium intra-distance (dashed line),
  $\delta d$ is the initial separate distance and $\Delta d$ the final bond-length; inset: changes
  in cell volume (black line) and atomic charges (red lines).}
  \label{fig:O-dimer}
\end{figure}

Accommodation oxygen dimer in UO$_{2}$ leads to a swollen of
the system volume (see table \ref{tab:coheE}). The induced stress forces
them orientated in $\langle111\rangle$ direction and occupy the Willis O$^{''}$ sites actually.
In energetics, oxygen dimer in UO$_{2}$ is meta-stable,
see figure \ref{fig:form-E}. Its decomposition process can be modeled by successively
moving the interstitial oxygen (as a test atom) in $C4_{1d}$ along the $\langle111\rangle$ direction
till to the cubic center, which is the most possible separate path. The resulted
potential shape
is shown in figure \ref{fig:O-dimer},
where $\delta \mathrm{d}$ is the initial depart distance between the two oxygens
and $\Delta \mathrm{d}$ the final (dimer) length.
The \emph{structure frozen} line was obtained by fixing the cell and all other atomic
positions whereas the \emph{optimized} one resulted from a fully relaxation of
the cell volume and shape and the nearby atomic coordinates that surrounding the defect.

Note a distance of
$\delta \mathrm{d}=2.2$\,$\mathrm{\AA}$ stands for the state that
the initial position of the test oxygen already close to the
cubic center. From figure \ref{fig:O-dimer} we get the critical distance to break a dimer is about
$1.73$\,\AA, with a barrier of $0.21$\,eV. Inset gives the variations of system
volume and Bader atomic charges of the two oxygens,
demonstrating a drastic behavior around
the breaking point. Two points need to be noticed here: the large charge transfers
and the contraction of system volume.
The later confirms that atomic size effect
is not an important factor for oxygen interstitials
in UO$_{2}$ where chemical interaction overwhelming. A deduction
of this is that a single
oxygen interstitial can occupy a site other than the cubic center, regardless it has the largest
space. Indeed, no experiment has detected the occupation of this site in
UO$_{2+x}$ when $x\geq 0.1$.
Chemical interaction might prefer other sites if volume is expanded. As the
``structure frozen'' line shows, interstitial oxygen is apt to forming dimers when
the volume is fixed at $651.49\,\mathrm{\AA}^{3}$.
Therefore, oxygen dimers may also exist at regions
with negative stresses.

As the limit case of an interstitialcy diffusion model, to form
an oxygen dimer in UO$_{2}$ requires an energy of $\sim 1.75\,$eV (figure \ref{fig:O-dimer}), compatible with
the NEB migration energy of $1.1\,$eV.\cite{durinck06} This
magnitude of migration energy corresponds to $\delta d\simeq 1.8\,\mathrm{\AA}$,
with an equilibrium intra-atomic distance about $2.0\,\mathrm{\AA}$
and the atomic charges $\sim -1.0\,e$. Therefore a charge oscillation
induced by oxygen diffusion is about $0.2\,e$, almost the same level as to oxidize
one uranium.

\subsection{Vibrational frequencies}
\label{sec:vibf}

Raman and infrared spectroscopies provide information about atomic
vibrations. These techniques can be employed to detect defect
clusters by searching the characteristic vibrational frequencies.
At finite temperatures, these frequencies directly contribute to
the formation energy and structural thermodynamic stability.

\begin{table}[]
\caption{\label{tab:freq} First principles results for structural, energetic,
and vibrational properties of oxygen interstitial and O$_{2}$ dimer in different
configurations. For comparison, calculated values for O$_{2}$ in vacuum are also listed.
$\Delta E$ is the energy difference between interstitial O/O$_{2}$ and vacuum O$_{2}$
(formation energy per pair interstitials), $d_{0}$ is the equilibrium bond length, $q$ is the Bader effective
charge, $\omega$ is the harmonic frequency. Note the $q$ in the last row is
just for to label the experimental condition.}
\begin{ruledtabular}
\begin{tabular}{l c c c c l} 
  Label & $\Delta E$\,(eV) & $d_{0}$\,(\AA) & $q$\,($e$) & $\omega$\,(cm$^{-1}$)  \\
\hline
  $C2_{1}$  & -4.496   & - &-1.15  & 292.5,\,316.7,\,403.9 \\
  $C4_{1}$  & -2.993   & - &-1.14  & 373.3,\,386.6,\,397.5\\
  $C4_{1d}$ & 0.159    & 1.47 & -0.61(-0.77) & 273.6,\,345.3,\,353.9 \\
            &          &      &              & 452.8,\,473.6,\,795.4 \\
  $C4_{2d}$ & -1.642   & 1.39  &-0.66,\,-0.59  & 447.4,\,482.3,\,496.6   \\
            &          &       &               & 608.5,\,637.2,\,995.4 \\
  Vacuum    &  0       & 1.22 & 0.0 & 1588.6   \\
  Expt.     &  - & 1.21/1.49\footnotemark[1] & 0.0/-2.0 & 1580.2\footnotemark[2]/-   \\
\end{tabular}
\footnotetext[1]{\, Ref. [\onlinecite{Jolly85}]}
\footnotetext[2]{\, Ref. [\onlinecite{huber79}]}\\
\end{ruledtabular}
\end{table}

Vibrational frequencies of single oxygen interstitial (has three modes) and dimer
(has six modes) in $C2_{1}$,
$C4_{1}$, $C4_{1d}$ and $C4_{2d}$ configurations were calculated.
In all calculations, we aligned the magnetic ordering direction along
the shortest axis (S), which always has the lowest energy.
Only harmonic frequencies were computed
here and has omitted all anharmonic effects. For fluorite UO$_{2}$, we have
checked that the contribution from the latter is
very small for oxygen and uranium interstitials (within $\pm 3\, \mathrm{cm}^{-1}$).
Table \ref{tab:freq} lists the calculated frequencies ($\omega$), as well as the equilibrium bond length
for dimers ($d_{0}$) and formation energies ($\Delta E$).
Due to the compression from oxygen cage, the vibrational
frequencies in $C4_{2d}$ have greater value than their counterparts in $C4_{1d}$. The stretch model
of O$_{2}$ molecule (with the largest $\omega$) has been greatly softened when incorporated in UO$_{2}$.
This is analogous to the incorporation of H$_{2}$ in an interstitial position of
semiconductors,\cite{walle98} where a decrease of the binding energy, an increase
in the bond length, and a lowering of the vibrational frequency was observed. The
underlying physics, however, might be different. In this case, by comparing the
calculated Bader effective charges with the partially ionic model of UO$_{2}$,\cite{yamada00} we can
identify the nominal charge of the oxygen dimers should be about $-2.0\,e$.
The variation of bond length confirms this interpretation. Consequently, the frequency
of stretch model is lowered from 1588.6\,$\mathrm{cm}^{-1}$ to 995.4\,$\mathrm{cm}^{-1}$
in $C4_{2d}$ and 795.4\,$\mathrm{cm}^{-1}$ in $C4_{1d}$.

\begin{figure}
  \includegraphics*[0.22in,0.19in][3.9in,2.96in]{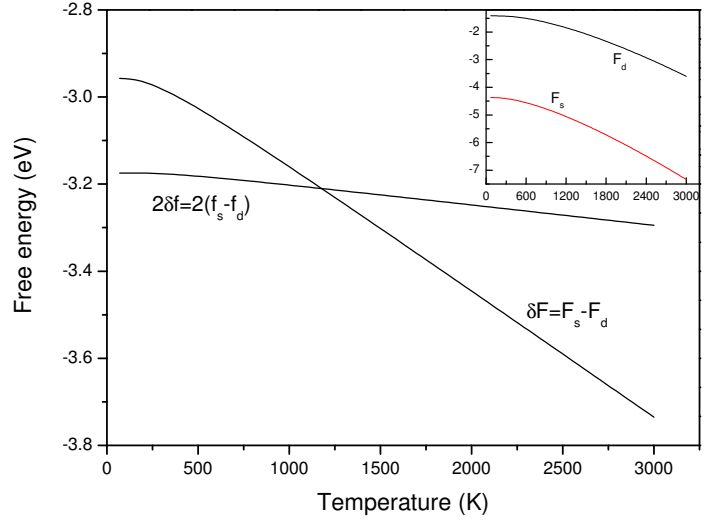}
  \caption{Variation of free energy difference contributed from interstitial vibrations.
  Inset: the free energies of a dimer in $C4_{2d}$ and its relative stable state $C2_{1}$ as
  a function of temperatures.
  }
  \label{fig:free-E}
\end{figure}

According to the calculated static energies, $C4_{2d}$ will decay to $C2_{1}$,
and $C4_{1d}$ to $C4_{1}$ eventually (see $\triangle E$ in table \ref{tab:freq}).
Computed frequencies indicate thermal vibrations would accelerate this process further.
Figure \ref{fig:free-E} gives the difference of free energy between $C4_{2d}$ ($F_{d}$),
$C2_{1}$ ($F_{s}$), $C4_{1d}$ ($f_{d}$) and $C4_{1}$ ($f_{s}$)
calculated with their formation energies (table \ref{tab:coheE}) and
vibrational frequencies (table \ref{tab:freq}), respectively.
The rapid drop of the free energy differences with increased temperature
implies that meta-stable oxygen dimers in UO$_{2}$ have a very short lifetime at finite
temperatures, and with little possibility to occupy
the cubic center sites: they must have been decomposed before enter the
oxygen cage.

\subsection{Defect clustering in UO$_{2+x}$}
\label{sec:clustering}

This section devotes to the possible defect \emph{clustering pattern} in UO$_{2+x}$.
Instead of compute the formation energies directly, we focus on the \emph{local
stability} of O$^{'}$ and O$^{''}$ sites in different circumstances here. This method cannot
determine what cluster is the most stable one, but it \emph{does} rule out
some combinations of O$^{'}$, O$^{''}$ and oxygen vacancies.

For this purpose we calculated the potential landscape
felt by a test oxygen atom. Just one fluorite cubic unit cell
was used. Here since the cell and all atomic coordinates have been frozen
up except that of the test oxygen, the error introduced by periodic
conditions is proportion to the second order of the charge density variation $\delta \rho$
that induced by images of the test atom.
This precision is enough for a qualitative analysis (Mind ionic interactions among the
test atom's images contribute only a constant to the energy and therefore
irrelevant to the problem).

\begin{figure}
  \includegraphics*[0.22in,0.19in][4.20in,3.05in]{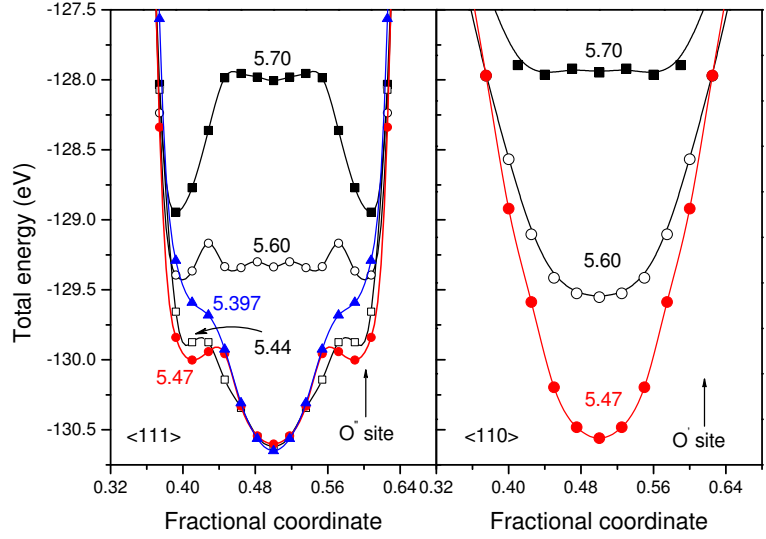}
  \caption{(Color online) Potentials for an oxygen interstitial in UO$_{2}$
  along $\langle111\rangle$ (left) and $\langle110\rangle$ (right) direction crossing the cubic center.
  The numerics refer to lattice constant and arrows point to the position of O$^{'}$
  or O$^{''}$ site. Fractional coordinate 0.5 denotes the cubic center.
  }
  \label{fig:pot-sp1}
\end{figure}

\subsubsection{Local stability of basic clustering modes}
\label{sec:local-stab}

At first we check the local stability of a single O$^{'}$ and O$^{''}$ site. Figure
\ref{fig:pot-sp1} shows the potential shapes crossing these two sites. It is seen
that O$^{''}$ site becomes meta-stable when the lattice constant increases to about $a=5.44$\,\AA.
And isotropic expansion stabilizes this site further which makes it the global
minimum if $a\geq5.6$\,\AA. A single O$^{''}$ interstitial actually forms a dimer with the
nearest lattice oxygen and this behavior
is in consistence with the structure frozen curve in figure \ref{fig:O-dimer}.
However, this effect does not benefit the stabilization of O$^{'}$ site.
Under ambient conditions the experimental lattice constant for UO$_{2+x}$ is within $5.45\sim5.47\,$\AA,
therefore a single O$^{'}$ or O$^{''}$ oxygen interstitial (as well as clusters formed by them only) is almost unstable.

\begin{figure}
  \includegraphics*[0.22in,0.19in][4.20in,3.05in]{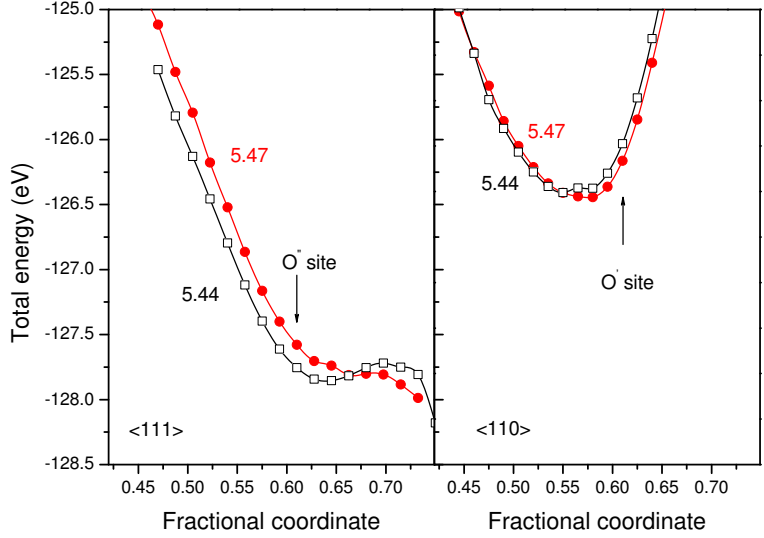}
  \caption{(Color online) Potentials for an oxygen interstitial in UO$_{2}$
  (incorporated with a V-O$^{'}$ pair) along $\langle111\rangle$ (left) and $\langle110\rangle$ (right) direction crossing the cubic center
  (fractional coordinate 0.5).
  }
  \label{fig:pot-sp2}
\end{figure}

The simplest cluster involved one oxygen vacancy, say a V-O$^{'}$ or V-O$^{''}$ pair,
is obviously unstable since nothing can prevent them from annihilation.
The next triple cluster is V-O$^{'}$(O$^{''}$) pair stabilized by an O$^{'}$ or O$^{''}$
interstitial. Considering the short distance between the nearest O$^{'}$ and O$^{''}$ sites,
the situation should be quite similar for them. Therefore hereinafter we only consider the
cases that incorporated with V-O$^{'}$ pairs. The potential shape
for an O$^{'}$(O$^{''}$)-V-O$^{'}$ cluster was calculated in an analogous
manner
except a lattice oxygen $(0.75,0.75,0.75)$ was moved to $(0.883,0.5,0.883)$, a nearest
O$^{'}$ site, to create the V-O$^{'}$ pair, as shown in figure \ref{fig:pot-sp2}.
Although the curve along $\langle110\rangle$ already changes asymmetrically about the cubic center (with
a fractional coordinate 0.5),
O$^{'}$ is locally unstable since it will decay to O$^{''}$ (with a swallow trap) then finally
to a position beyond the $(0.75,0.75,0.75)$ site. This rules out the O$^{'}$-V-O$^{'}$ (V-2O$^{'}$)
and O$^{''}$-V-O$^{''}$ (V-2O$^{''}$) triple clusters that distribute symmetrically about a lateral of the oxygen
cage.

\begin{figure}
  \includegraphics*[0.22in,0.19in][4.20in,3.05in]{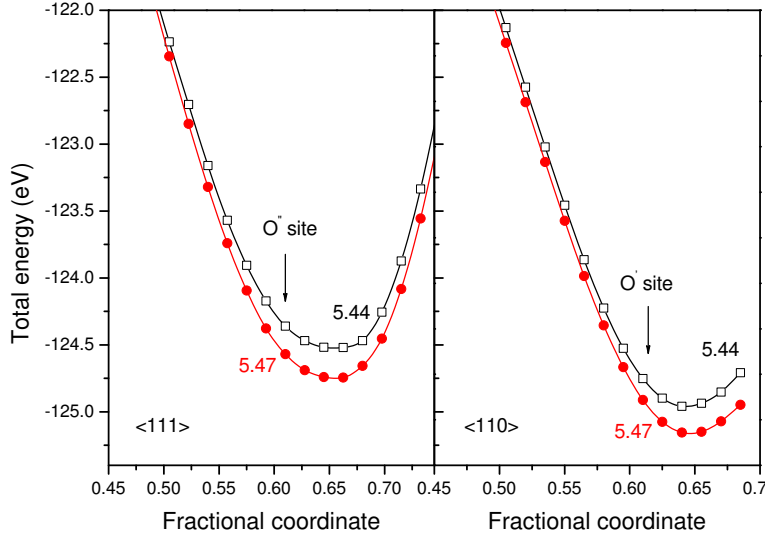}
  \caption{(Color online) Potentials for an oxygen interstitial in UO$_{2}$
  (incorporated with two V-O$^{'}$ pairs)
  along $\langle111\rangle$ (left) and $\langle110\rangle$ (right) direction crossing the cubic center
  (fractional coordinate 0.5).
  }
  \label{fig:pot-sp3}
\end{figure}

To stabilize the O$^{'}$ site locally, we have tried all possible combinations
and find two nearest oxygen vacancies seems necessary. Figure
\ref{fig:pot-sp3} gives the potentials that incorporated with two V-O$^{'}$ pairs. Mind these O$^{'}$ sites
should be in otherwise empty oxygen cubes that do not share the lateral linking the two
vacancies with the original one. The pairs are thus created by moving $(0.75,0.75,
0.75)$ oxygen to $(0.617,1.0,0.883)$ and $(0.75,0.75,0.25)$ to $(1.0,0.617,0.117)$,
respectively. We see it do prefer the O$^{'}$ but not O$^{''}$ site. In fact this cluster
would become the Willis 1:2:2 (O$^{'}$:V:O$^{''}$) cluster\cite{willis64b} if move the two O$^{'}$ interstitials to their nearest
O$^{''}$ sites and form two V-O$^{''}$ pairs rather than the V-O$^{'}$ ones. Figure \ref{fig:pot-sp3} shows it might be
local stable, in consistent with empirical calculations.\cite{catlow77}
The stabilization of O$^{''}$ by V-O$^{'}$ pairs sited in the otherwise empty oxygen cages is unclear
by figure \ref{fig:pot-sp2}, but calculations show a V-2O$^{'}$ triple do stabilize
O$^{''}$ locally (O$^{''}$-V-2O$^{'}$), as well as a V-2O$^{''}$ triple can (O$^{''}$-V-2O$^{''}$).

Thus we finally arrive at the conclusions that: ({\romannumeral 1}) O$^{'}$ or O$^{''}$ interstitials
cannot exist by themselves;
({\romannumeral 2}) each O$^{'}$ site must incorporate with two nearest oxygen
vacancies, while O$^{''}$ can be stabilized by a V-2O$^{'}$(O$^{''}$) triple.
That means the possible clustering pattern for oxygen defects should only be:
(\emph{a}) V-3O$^{''}$ or V-4O$^{''}$ isolated clusters, in the same manner
of split-interstitial where several atoms sharing a single lattice site; (\emph{b}) cluster chains of
V-O$^{'}$-V or V-2O$^{'}$-V by sharing the vacancy sites. These chains should be closed to have
all O$^{'}$ interstitials locally stable whilst minimize the vacancy/interstitial ratio;
(\emph{c}) cluster of V-(2)O$^{'}$-V chains terminated by two V-(2)O$^{''}$ clusters
at both of the extreme sides by sharing the vacancy sites. We call these small fractal clusters the \emph{Willis type} cluster,
including 1:2:2,\cite{willis64b} 2:2:2,\cite{willis78,allen82} 4:3:2\cite{catlow77,cheetham71} clusters and so on.
However, their actual stability is still unknown which requires accurate knowledge
about their formation free energies.

\subsubsection{Phase diagram for clusters}

In UO$_{2+x}$, the positive formation energy of oxygen Frenkel pair and
the small energy gain from interactions among interstitials (see figure \ref{fig:form-E})
implies the only way to reduce the energy increment from creating vacancies is via
vacancy-interstitial (V-I) interactions. Obviously the nearest V-I pairs contribute
the most. Therefore the relative stability of clusters can be judged
roughly by counting the number of nearest interstitials around each vacancy. For
example in a 1:2:2 cluster, each vacancy has only two V-I pairs, while in 2:2:2 it has
three and in 4:3:2 it is 3.3. That means 1:2:2 should be meta-stable, even though
it can explain the concentrations measured by Willis in 1964.\cite{willis64b}
However, this data also can be explained by a larger cluster with 4 O$^{''}$
interstitials, namely, a 2:2:4 cluster where each vacancy has four V-I pairs.
Willis type clusters are necessary in order to explain the large
concentration of O$^{''}$ interstitials, which is impossible by cuboctahedral
cluster only (belonging to pattern (\emph{b})). For example the data for crystal A done by Murray \emph{et al}.\cite{murray90} obviously
belongs to 2:2:2 clusters while crystal B should be a mixture of 4:3:2 and
cuboctahedral clusters or a 6:4:2 cluster.

However, a big Willis type cluster is unfavorable since the disturbance to
fluorite lattice is proportion to its size linearly.
A similar situation holds for a loosely closed chain of the pattern (\emph{b}).
In this sense
the most regular and close-packed defect cluster, the cuboctahedral cluster,
takes the advantage of sharing the space
with all vacancies and interstitials to minimize the damage to the matrix.
Also, one fluorite cubic cell can accommodate one (or less) Willis type cluster
or one cuboctahedral cluster. But the former provides only 2 excess anions
while the latter provides 5. When composition $x$ increased,
cuboctahedral cluster has a big advantage over Willis type cluster,
not to mention each of its vacancy has a number of V-I pairs greater
than three. As for the clustering pattern (\emph{a}), though there are 3 (V-3O$^{''}$)
or 4 (V-4O$^{''}$) V-I pairs for each vacancy, we can discard them at first since
no experiments showed so high concentration for O$^{''}$ interstitials.

\begin{figure}
  \includegraphics*[]{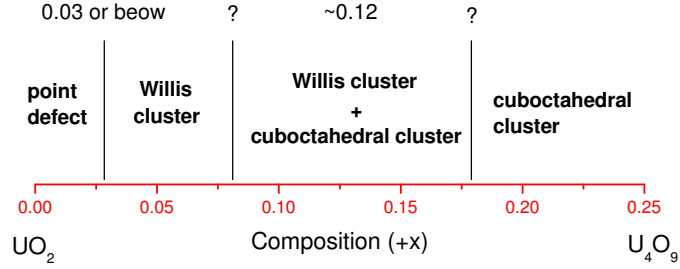}
  \caption{(Color online) A schematic \emph{phase diagram} for oxygen defect clustering in UO$_{2+x}$.
  The boundaries, however, are not clearly known.
  }
  \label{fig:cluster-pd}
\end{figure}

It becomes evident when check the variation of O$^{'}$:O$^{''}$ ratio as $x$ increased:
around $x=0.11\sim0.13$, three different data sets were observed (0.08:0.16,\cite{willis64b}
0.14:0.12, and 0.33:0.10\cite{murray90}), implying the occurrence of Willis type
clusters. As $x$ approaches to 0.25, however, this ratio increases drastically\cite{bevan86,cooper04,garrido03}
and shows the dominance of cuboctahedral clusters. Therefore, by taking
the stability of point interstitial at low $x$ into account, one concludes that there is a quasi- \emph{phase
diagram} for oxygen clusters in UO$_{2+x}$, as shown in figure \ref{fig:cluster-pd}.\cite{note3}
To determine the exact geometry of ground clusters and their boundaries
would be the center of future works in this field.

It is worthwhile to point out that such kind of defect clustering is not
unique to uranium dioxide. According to the formation energy of point defects,
one can classify binary compounds into three classes: (\emph{A}) all formation
energies are positive, (\emph{B}) only one of the formation energies is negative,
and (\emph{C}) both cation interstitial (vacancy) and anion vacancy (interstitial)
have negative formation energies. There is no off-stoichiometry driven force in case (\emph{A}),
and disfavors extensive defect clustering. However the negative formation energies
in the other two cases will drive the system to a non-stoichiometric composition
where defect clustering becomes favorable. This is because
the interaction among defects can lower the system energy greatly, and lead to a pure defect
clustering (via a full vacancy-interstitial annihilation) or mixed defect clusters
that contain both vacancy and interstitial.
Also,
the mixed cluster is possible only
when the point defect with positive formation energy (vacancy or interstitial) has
the function to stabilize the other defects in an energy favorable configuration
(in a similar concept of the split-interstitial defect mechanism).
Obviously UO$_{2}$ fulfills these conditions (see figure \ref{fig:form-E} and
previous subsection) and belongs to case (\emph{B}), where oxygen interstitial has
a negative formation energy and clustering involves no uranium
sites. On the other hand, case (\emph{C}) contains clusters composed of both cation and anion defects,
and might exhibit more complex behaviors.

\subsection{Concentration of defects}
\label{sec:concentration}
\subsubsection{Generalization of the PDM}

The point defect model (PDM) was introduced by Matzke\cite{matzke87}
and Lidiard\cite{lidiard66} to analyze the populations
of defects in UO$_{2+x}$, where $x$ indicates the deviation from stoichiometry.
This model is based on the hypothesis that the defects responsible for
the deviation from stoichiometry in UO$_{2+x}$ are isolated point defects.
However, it has been known for long that oxygen interstitials form clusters,
and usually PDM performs poorly even at small $|x|$.\cite{crocombette01,freyss05} Therefore
it is worthwhile to generalize this model beyond the point approximation.
Since traditionally defect concentrations are defined in a lattice model
as the number of defects present divided by the number of available sites
for the defect under consideration,
the most general and elegant generalization of PDM would be cluster
variation method (CVM),\cite{kikuchi51} which also bases on lattice gas model and
computes cluster configurational entropy
explicitly. The related effective cluster interactions can be determined
by cluster expansion method (CEM).\cite{connolly83}
For UO$_{2+x}$, at first sight it seems being a quaternary system
($V_{O}$, $V_{U}$, $I_{O}$, and $I_{U}$) and cannot be
tackled by modern CVM and CEM techniques. But since defects on the uranium
subsystem usually are isolated point defects that couple
with oxygen subsystem via Schottky defects,
in fact only oxygen defects need to be treated explicitly.
However, in order to include O$^{'}$ and O$^{''}$ sites in the calculation, one has to use an
extended lattice, which introduces another two difficulties.

The first one relates to the local stability of O$^{'}$ and O$^{''}$ sites, since these sites
are not well defined and usually a full relaxation is required to get
the optimized geometry. However, in most configurations they are not at
the potential minimums and makes it impossible to include
the relaxation effects in the \emph{ab initio} CEM procedure. Fortunately, an algorithm
proposed by Geng \emph{et al.} can tackle this problem simply.\cite{geng06}
The second difficulty is that most configurations on the extended
oxygen sublattice is unphysical, \emph{i.e}., some distances among oxygen sites
are too short to allowed. To exclude these unphysical configurations,
one has to use \emph{loose clusters} to expand the energy, which
deteriorates the convergence of cluster expansion drastically.

If all non-negligible clusters are independent and uncorrelated, a simple approximation
exists to calculate cluster populations.
Two clusters are called independent if none of them is the other one's sub-cluster (or loosely,
cannot dissociate or combine into other clusters).
This ensures all cluster concentrations are completely independent.
Assume there are $M$ such kind of clusters under consideration, then the system free energy can
be written as
\begin{equation}
  F=\sum_{i=1}^{M}\rho_{i}\left(E_{i}+\kappa_{B} T \ln \rho_{i}\right)
\end{equation}
in the closed regime (in which the system cannot exchange atoms with
the exterior). Here $E_{i}$ stands for the $i$-th cluster's formation
energy.
Minimization this free energy with respect to each cluster density $\rho_{i}$
(under the condition that $x$ is fixed)
gives
\begin{equation}
  \rho_{i}=g_{i}\exp\left(\frac{-E_{i}}{\kappa_{B} T}\right),
  \label{eq:rho}
\end{equation}
associated with the composition equation
\begin{equation}
  x=f(\rho_{1},\cdots,\rho_{M}).
\end{equation}
In Eq.(\ref{eq:rho}) the factor $g_{i}$ is introduced to account for the degeneracy
if the cluster has internal freedom, while non-degenerated states can be treated as independent.
This gives the internal entropy contributions and is the most significant difference
between the independent clusters approximation (ICA) and PDM.

The PDM equations can be derived by considering only isolated point
defect excitations (without internal structure): $V_{O}$, $V_{U}$, $I_{O}$ and $I_{U}$.
In a closed system the particle numbers must be conserved, reduces
the number of independent defect modes to three. On the other hand,
the formation energy reference state for point oxygen and uranium defects
usually are different, therefore one should instead use three independent combinations of
these isolated defects to eliminate this ambiguity. The
simplest candidates are oxygen and uranium Frenkel pairs and Schottky defect (or
equivalently, anti-Schottky defect). Consequently, $M=3$ and $i=1,\cdots,3$
corresponds to the isolated Frenkel pairs and Schottky defect, respectively.
In this way Eq.(\ref{eq:rho}) becomes
\begin{align}
  &\rho_{FP_{O}}=\exp\left(\frac{-E_{FP_{O}}}{\kappa_{B} T}\right)\equiv [V_{O}][I_{O}]\label{eq:PDM1},\\
  &\rho_{FP_{U}}=\exp\left(\frac{-E_{FP_{U}}}{\kappa_{B} T}\right)\equiv [V_{U}][I_{U}]\label{eq:PDM2},\\
  &\rho_{S}=\exp\left(\frac{-E_{S}}{\kappa_{B} T}\right)\equiv [V_{O}]^{2}[V_{U}]\label{eq:PDM3},
\end{align}
and the composition equation expressed in point defect populations
\begin{equation}
  x=2\left([V_{U}]-[I_{U}]\right)+[I_{O}]-2[V_{O}].
  \label{eq:PDM4}
\end{equation}
Eqs.(\ref{eq:PDM1}$\sim$\ref{eq:PDM4}) comprises the PDM equations exactly.

To include cluster effects, taking that in $C4_{2}$ configuration as example, we need reinterpret
the two interstitials as an \emph{isolated} diagonal pair (\emph{dp}). Assuming
this interstitial pair is predominant over the point one,
then Eq.(\ref{eq:PDM1}) is replaced by
\begin{equation}
\rho_{dp}[V_{O}]^{2}=\exp\left(\frac{-E_{dp}-2E_{V_{O}}}{\kappa_{B} T}\right),
\label{eq:pd1}
\end{equation}
where two isolated oxygen vacancies have been introduced to eliminate the ambiguity
in extrinsic defect formation energy. Also, the composition equation becomes
\begin{equation}
  x=2\left([V_{U}]-[I_{U}]\right)+\rho_{dp}-2[V_{O}],
  \label{eq:conecntra2}
\end{equation}
with Eqs.(\ref{eq:PDM2}) and (\ref{eq:PDM3}) keep unchanged. This procedure can be extended
to include other independent clusters easily.

\subsubsection{Defect concentrations in PDM}

\begin{table}[]
\caption{\label{tab:FE-FS} Formation energies (eV) of point defects in UO$_{2}$: uranium
and oxygen vacancies (U-Vac and O-Vac), uranium and oxygen interstitials (U-Int
and O-Int), Frenkel pairs (O-FP and U-FP), and Schottky defect (S).}
\begin{ruledtabular}
\begin{tabular}{l r r r l c r l} 
   & U-Vac & O-Vac & U-Int & O-Int & O-FP & U-FP & S \\
\hline
  LSDA+U\footnotemark[1] &9.1 & 7.5&8.2 & $-$2.2& 5.4&17.2 &10.6\\
  GGA+U\footnotemark[2] &8.4 & 4.5&4.7 & $-$0.4& 4.0&13.1&5.8\\
  GGA\footnotemark[3]  & 4.8   & 6.1 &7.0  &$-$2.5&3.6&11.8&5.6  \\
  GGA\footnotemark[4]  & 5.1   & 6.1 &7.5  &$-$2.6 &3.5&12.6&6.0 \\
  LDA\footnotemark[5] & 3.3    & 6.7 & 7.3 & $-$2.9 &3.9&10.7&5.8 \\
  LDA-LMTO\cite{petit98} & 19.1   & 10.0  &11.5&$-$3.3 &6.7&30.6&17.1  \\
  Semi-empirical\cite{jackson87}    &  80.2&16.9 &$-$60.8 &$-$12.1 & 4.8&19.4&11.3  \\
  PDM estimates\cite{matzke87}       &  $-$ & $-$ & $-$    & $-$ &3.0$\sim$5.8 & 9.5 & 6.0$\sim$7.0   \\
\end{tabular}
\footnotetext[1]{\, this work, with 8 fluorite cubic cells}
\footnotetext[2]{\, with 8 fluorite cubic cells, Ref. [\onlinecite{iwasawa06}] }
\footnotetext[3]{\, with 2 fluorite cubic cells, Ref. [\onlinecite{freyss05}]}
\footnotetext[4]{\, with 1 fluorite cubic cell, Ref. [\onlinecite{freyss05}]}
\footnotetext[5]{\, with 2 fluorite cubic cells, Ref. [\onlinecite{crocombette01}]}\\
\end{ruledtabular}
\end{table}

\begin{figure}
  \includegraphics*[]{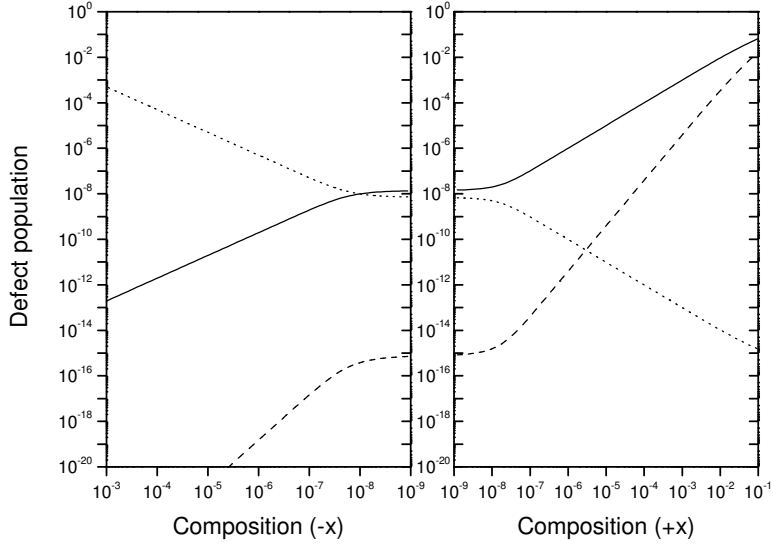}
  \caption{Analysis of the point defect model at a temperature of 1700\,K. Variation of the concentrations
  of point defects with the deviation from stoichiometry $x$: hypostoichiometric regime (on the left) and
  hyperstoichiometric regime (on the right). Solid (respectively dotted and
  dashed) line indicate the concentration in oxygen interstitial (respectively
  oxygen vacancy and uranium vacancy). The concentration of uranium interstitial is negligible.
  }
  \label{fig:conc-pdm}
\end{figure}

In the poind defect approximation, the formation energy of a Frenkel pair
of the $X$ specie is given by
\begin{equation}
  E_{FP_{X}}=E^{N-1}_{V_{X}}+E^{N+1}_{I_{X}}-2E^{N}
\end{equation}
and for the Schottky defect (\emph{S}) by
\begin{equation}
  E_{S}=E^{N-1}_{V_{U}}+2E^{N-1}_{V_{O}}-3\frac{N-1}{N}E^{N},
\end{equation}
with $N$ the number of atoms and $E^{N}$ the total (or cohesive) energy in the defect-free supercell;
$E^{N\pm1}_{V_{X},I_{X}}$ the energy of the cell with the defect. Here we use
$C8_{\pm1}$ and $^{u}C8_{\pm1}$ to model the point defects, thus $N=96$
and $E^{N}$ and $E^{N\pm1}_{V_{X},I_{X}}$ can be obtained by timing 8 to the corresponding
cohesive energies listed in table \ref{tab:coheE}.

The formation energies of the defects obtained are listed in table \ref{tab:FE-FS}.
They are compared to the previous theoretical results \cite{crocombette01,petit98,freyss05,jackson87,iwasawa06}
and PDM estimates based on diffusion measurements.\cite{matzke87}
Note that the GGA+U employed the same $U$ parameter as in this work. A detailed comparison of its
results with LSDA+U
can be found in Ref.[\onlinecite{geng07}]. Despite it produced a similar band gap
and local magnetic moment as LSDA+U, it predicted a \emph{big} lattice constant $\sim5.55$\,\AA.
By figure \ref{fig:pot-sp1} we know this would lead to an underestimation
of the oxygen interaction with the matrix. Table \ref{tab:FE-FS} proves this
by showing a smaller absolute value of the oxygen interstitial and vacancy formation
energies than any other calculations. However, this failure is not from GGA
but the parameter of $U$.\cite{note2} Besides, this $U$ also underestimates the formation
energy of uranium interstitial greatly, implying one needs to fitting an own
$U$ value for GGA functional separately.

The improvement of LSDA+U over the pure GGA or LDA results is significant.
Both the latter underestimate the formation energy of uranium vacancy by about 2 times, and $10\%\sim20\%$
for oxygen vacancy. By the lump, LSDA+U corrects the energy for O-FP by
a $50\%$ and $38\%$, a $46\%$ and $61\%$ for U-FP, an $89\%$ and $83\%$ for
Schottky defect over GGA and LDA, respectively. This correctness makes our
LSDA+U results the first \emph{ab initio} defect formation energies
that \emph{predict} the predominance of oxygen defects within a broad enough stoichiometric
range over uranium ones (For the performances of LDA or GGA formation energies and
the PDM anticipation, please see Refs.[\onlinecite{crocombette01,freyss05}]).

The defect concentrations, or equivalently populations, calculated with PDM equations are shown in
figure \ref{fig:conc-pdm}. An arbitrary temperature of 1700\,K is chosen.
 We see oxygen interstitial dominates when $x>0$
while it is oxygen vacancy when $x<0$. At $x\sim0$, O-FP overwhelms.
This picture is in a good agreement with diffusion measurements \emph{interpreted}
by PDM,\cite{matzke87} but different from neutron diffraction data where
non-negligible oxygen vacancies were observed when $x>0$.\cite{willis64a,willis64b,willis78,bevan86,murray90,cooper04}
The population of oxygen vacancy predicted by PDM is too low to be true. To increase this
population in the regime $x>0$, one needs to take clustering effect into account.

\subsubsection{Defect concentrations with independent clusters approximation}

\begin{figure}
  \includegraphics*[]{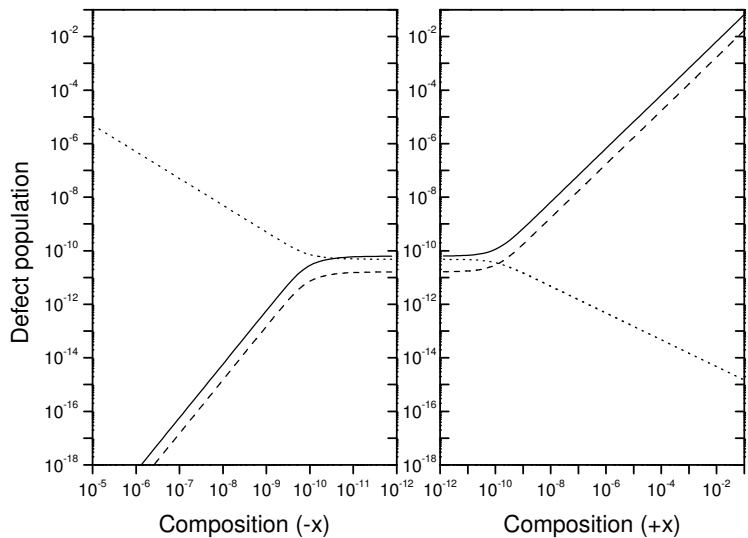}
  \caption{Analysis of independent clusters model with isolated diagonal oxygen interstitial pair
  approximation. Others are the same as in figure \ref{fig:conc-pdm}.
  }
  \label{fig:conc-ica}
\end{figure}

Assume the oxygen diagonal pair in $C4_{2}$ is dominant over the single interstitial,
one can calculate the clustering effect formally. It is not a promising
assumption due to the small energy difference between them. While it can be
used to analyze the influence of \emph{pure} interstitial clusters that occupied
only the octahedral sites on the vacancy populations (they should have similar
effects). Also it serves to show how
the independent clusters approximation works out.

Using the defect formation energy of $C4_{2}$ and Eqs.(\ref{eq:PDM2}, \ref{eq:PDM3}, \ref{eq:pd1} and \ref{eq:conecntra2}),
we calculated the defect populations following the same manner as PDM, the result is
presented in figure \ref{fig:conc-ica}. Note here that $\rho_{pd}$ turns out to have
the same numeric value as $[I_{O}]$. We see that this pure clustering mechanism do decrease
the oxygen interstitial population, but that of oxygen vacancy in the $x>0$ regime is still too
low. Another problem raised here is the population of uranium vacancy is closely pinned
on that of the oxygen interstitial. It is not what we wanted. Roughly, figure
\ref{fig:conc-ica} suggests clusters associated with oxygen \emph{vacancies}
is necessary in order to enhance the latter's concentration greatly in $x>0$ regime and to pin on that of oxygen interstitial,
as implied by the neutron diffraction measurements.

\section{CONCLUSION}
\label{sec:conclu}

In summary, we performed a comprehensive calculation on defect properties in
UO$_{2\pm x}$ with LSDA+U method. The volume changes induced by defects and their
formation energies were computed accurately. Analysis of these energies for a
series configurations concluded that defect clustering is unavoidable when $x\geq0.03$,
compatible with experimental fact. Atomic charge calculations in Bader's definition, however, showed
the difficulty to oxidize uranium to U$^{6+}$ and the charged oxygen apt to
losing its electrons, against common expectation. As the simplest interstitial
cluster, oxygen dimer behaviors in a manner similar to a normal oxygen in energetics and
charge state. It was identified as ionic dioxygen molecule with two excess
electrons. Static and vibrational free energy calculations, however, showed it
quite unstable and might only be a transient state during oxidization process.

Oxygen dimer is the extreme case for interstitialcy diffusion of oxygen, which
may induce a charge fluctuation with a magnitude less than $0.2\,e$. It also
presents as a special case for Willis O$^{''}$ site occupancy under stretch. The
stabilization mechanism for this site under ambient conditions, however, is attributed to
a V-2O$^{''}$(O$^{'}$) triple by the local stability analysis.
Also, O$^{'}$ site is stabilized only by the nearest oxygen vacancy pair. This comprises
the basic clustering pattern for defects in UO$_{2+x}$: play with the
four building blocks (V-(2)O$^{''}$ and V-(2)O$^{'}$-V) by sharing the vacancy sites.
The actual stability of clusters should be judged by the formation
energies, which beyond the scope of this paper and we prefer to future work.
A quasi phase diagram for defect clusters vs composition was
also proposed to explain the observed population ratios of O$^{'}$ and O$^{''}$ sites,
which of course requires further refinement step by step when more calculations
and experimental data are available.

The formation energy of Frenkel pairs and
Schottky defect calculated with LSDA+U have been improved more than $50\%$ over the
GGA and LDA results. With these energies and the point defect model, we first time
showed the predominance of oxygen defects by first principles. Finally we generalized the PDM
to independent clusters approximation that allows us to compute the population of clusters,
and revealed the necessity to move
on to Willis type clusters.

\begin{acknowledgments}
This study was financially supported by the Budget for
Nuclear Research of the Ministry of Education, Culture, Sports,
Science and Technology of Japan, based on the screening and counseling by the
Atomic Energy Commission.
\end{acknowledgments}



\begin{thebibliography}{99}
\bibitem{note0} Here the logic goes as that
since the ordered phase $C4_{2}$ had not been observed
in experiments but defect clusters instead,
the unstability of random distrubited
non-interacting point oxygen interstitials against the $C4_{2}$ phase directly
supports the preference of defect clusters over the isolated
interstitials.
\bibitem{note1} There may have other complex mechanism
for oxygen adsorption and surface reconstruction that might modify the potential surface
and then the numerical values considerably, but this picture still holds
for deep-layers in a qualitative sense. This argument is also supported by Ferguson
and McConnell's report of about $-230$\,kJ mol$^{-1}$ ($\sim$ $-2.38$\,eV) heat is released out
in the chemisorption of molecular oxygen onto the surface of UO$_{2}$ during the
initial stages of the reaction, as quoted in Ref.[\onlinecite{McEachern98}], being comparable to the energy that would be released
if decompose an oxygen dimer in UO$_{2}$ matrix.
\bibitem{note3} Here no temperature influence is considered. A more realistic
description should include this effect since usually temperature will alter
cluster populations greatly.
\end{thebibliography}
\end{document}